\documentclass[sigconf,anonymous=false]{acmart}
\renewcommand\footnotetextcopyrightpermission[1]{} 
\usepackage{graphicx}
\usepackage{amsmath}
\usepackage{booktabs}
\usepackage{algorithm}
\usepackage{algorithmic}
\usepackage{amsfonts}
\usepackage{multirow}[c]
\usepackage{makecell}
\usepackage{subfigure}
\usepackage{color}
\usepackage{bm}
\usepackage{epstopdf}
\usepackage{url}
\usepackage[cal=cm]{mathalfa}
\usepackage{balance}
\usepackage{threeparttable}
\usepackage{wrapfig}
\usepackage{tabularx}
\usepackage{array}
\usepackage{enumitem}
\usepackage{appendix}
\usepackage{tikz}
\usepackage{pifont}

\setlist[itemize]{leftmargin=*}

\makeatletter
\newcommand\notsotiny{\@setfontsize\notsotiny\@vipt\@viipt}
\makeatother

\AtBeginDocument{%
  }

\begin{document}


\title{LazFormer: Scaling Transformers for Industrial Recommendation via Transferable Generative Pre-training}

\author{Xiaodong Li, Alin Fan, Mingyang Li, Yan Xiao, Shichao Nie, Junfeng Zhang, Shaochuan Lin, Zhanming Ou, Tao Luo, Xiaoyi Zeng}
\affiliation{
\institution{Alibaba International Digital Commerce Group, Beijing, China}
\country{\{limingxi.lxd, alin.fal, lmy398032, yanwei.xy, nsc383441, sichu.zjf, lin.lsc, zhanming.ozm,\\ luotao.lt, yuanhan\}@alibaba-inc.com}
}

\renewcommand{\shorttitle}{LazFormer: Scaling Transformers for Industrial Recommendation via Transferable Generative Pre-training}
\renewcommand{\shortauthors}{Xiaodong Li et al.}

\begin{abstract}


Transformers have shown promising performance in LLMs due to their outstanding scalability, 
while several studies have investigated the scalability of Transformers for industrial recommendation. They typically rely on a single ranking model to optimize both sparse and dense parameters from scratch, resulting in substantial computational resource consumption and slow convergence. 
Fortunately, the pre-training models offer an effective solution to the above issues by providing favorable initialization of both sparse and dense parameters for the subsequent ranking.
However, the pre-training and ranking paradigm still faces two major limitations:
(1) Since the input features used in pre-training and ranking are usually inconsistent, directly transferring dense parameters from pre-training to ranking may lead to negative transfer.
(2) Multi-epoch training during the ranking process may result in the overfitting of sparse parameters, while freezing the sparse parameters limits their adaptability to the ranking objectives.
To this end, we propose a
\textit{Scaling Transformer for Industrial Recommendation via Transferable Generative Pre-training},
termed \textbf{LazFormer}. Specifically, we first present a generative pre-training module to autoregressively generate sequential features, providing favorable initialization of both sparse and dense parameters for the subsequent ranking. 
To solve the negative transfer of dense parameters,
we propose a transferable residual adapter that injects additional ranking-specific features into ranking in a residual manner. Moreover, a request-aware ranking module integrates long-sequence compression, hybrid sparse attention, and a request-aware paradigm to efficiently model users' long sequences.
Besides, we further propose an asymmetric multi-epoch training strategy that resets sparse parameters while continuously accumulating dense parameters across epochs, alleviating the overfitting of sparse parameters. 
Extensive offline experiments and online A/B testing demonstrate the effectiveness and scalability of LazFormer. LazFormer has been deployed in an industrial recommendation system, 
delivering significant gains for the platform.

\end{abstract}



\maketitle

\section{Introduction}

The Transformer architecture has demonstrated remarkable success in large language models (LLMs)~\cite{qwen,qwen3,deepseek}, where model quality continuously improves by scaling model capacity, training data, and computational resources. The unified token-based architecture and outstanding scalability of Transformers have also inspired a paradigm shift in industrial recommendation. Traditional industrial recommendation models~\cite{deepfm,dien,dcn} usually follow an explicit feature interaction paradigm, in which high-dimensional sparse embeddings are processed by dedicated sequence modeling and feature interaction modules. 
More recently, Wukong~\cite{wukong} and RankMixer~\cite{rankmixer,tokenmixer_large} explore scaling feature interaction networks by increasing model parameters and computational budgets.
Nevertheless, industrial recommendation introduces unique challenges absent in LLMs, including billions of sparse features, rapidly evolving user interests, and strict latency constraints. 
Thus, how to fully unleash the potential of Transformers in industrial recommendation remains largely underexplored.

Recently, several studies~\cite{hstu,mtgr} have investigated the scalability of Transformer-based recommenders. Specifically, HSTU~\cite{hstu} and MTGR~\cite{mtgr} reformulate recommendation as a generative sequential transduction problem and demonstrate an empirical scaling law by increasing model capacity, sequence length, and training compute. More recently, OneTrans~\cite{onetrans} employs a unified Transformer backbone to jointly conduct user sequence modeling and feature interaction in a per-token manner, and subsequent studies~\cite{hyformer,mixformer} further extend this direction with heterogeneous feature interaction and long-sequence modeling. These methods demonstrate that Transformer-based recommenders can continuously benefit from scaling depth, width, and sequence length. 

Despite their encouraging progress, existing Transformer-based methods typically rely on a single ranking model to optimize both sparse and dense parameters from scratch, leading to substantial computational resource consumption and slow convergence.
Fortunately, pre-training models could provide favorable initialization of both sparse and dense parameters for the subsequent ranking process, offering an effective solution to the above issues.
However, the pre-training and ranking paradigm still faces two major limitations:
(1) \textbf{Negative transfer of dense parameters}. The features available during pre-training are usually inconsistent with those required by the ranking process. 
Directly incorporating these additional ranking-specific features into ranking may disturb the representation space expected by the pre-trained dense parameters, thereby leading to negative transfer. (2) \textbf{Overfitting of sparse parameters}. 
Continuously optimizing billions of sparse parameters for multiple epochs may lead to overfitting~\cite{meda,meda_v2} and destroy their transferable representations. Conversely, freezing the sparse parameters limits their adaptability to the ranking objectives. Thus, effectively transferring both sparse and dense parameters while co-scaling model capacity and effective training data consumption remains a critical challenge.

To address the above limitations, we propose \textbf{LazFormer}, a \textit{Scaling Transformer for Industrial Recommendation via Transferable Generative Pre-training}.
Specifically, LazFormer first performs generative pre-training over users' historical interaction sequences via autoregressive next-item prediction, jointly learning sparse and dense parameters to initialize subsequent ranking.
To mitigate the negative transfer of dense parameters, we introduce a transferable residual adapter that injects additional ranking-specific features into ranking in a residual manner. 
Moreover, we propose a request-aware ranking module for long-sequence modeling, integrating long-sequence compression, hybrid sparse attention, and a request-aware paradigm.
Besides, we introduce an asymmetric multi-epoch training strategy that resets sparse parameters 
while continuously accumulating dense parameters across epochs, enabling the Transformer backbone to absorb more ranking supervision while avoiding sparse overfitting.
Ultimately, these components enable the effective transfer of both sparse and dense parameters, as well as the co-scaling of model capacity and training data in LazFormer.

Our main contributions can be summarized as follows:
\begin{itemize}
    \item We propose LazFormer, a scaling transformer for industrial recommendation via transferable generative pre-training, which jointly transfers sparse and dense parameters from generative pre-training to downstream ranking while alleviating dense negative transfer and sparse overfitting.

    \item We introduce a transferable residual adapter for stable sparse and dense transfer, a request-aware ranking module for efficient model scaling, and an asymmetric multi-epoch training strategy for effective training data scaling.

    \item We conduct extensive offline experiments and online A/B testing to validate the effectiveness of LazFormer. LazFormer has been deployed in a real-world industrial recommendation system, delivering significant gains for the platform.
\end{itemize}



\section{Problem Definition}

In industrial recommendation, we focus on predicting multiple probabilities that reflect a user's preference for a target (candidate) item, including: the click-through rate (\textit{i.e.}, \textbf{CTR}) and the post-click conversion rate (\textit{i.e.}, \textbf{CVR}).

Formally, we use $\mathcal{D}$ to denote the user-item interaction data, which can be formulated as $\mathcal{D}=\{\mathcal{U},\mathcal{V},\mathcal{Y}\}$. Specifically, $\mathcal{U}$, $\mathcal{V}$ and $\mathcal{Y}$ represent the user set, item set and several true labels (\textit{i.e.}, $y^{ctr}\in\{0,1\}$, and $y^{cvr}\in\{0,1\}$), respectively. For instance, $y^{ctr}_{ij}\in\{0,1\}$ denotes whether user $u_i$ clicks item $v_j$. Moreover, the historical interaction sequence of user $u_i$ can be denoted as $\mathcal{H}_i=\{v_j|v_j\in\mathcal{V}\}_{j=1}^{|\mathcal{H}_i|}$, where each $v_j$ is associated with multiple distinct categories of features, 
including ID-based features and Side Info-based features:
\begin{itemize}
    \item \textbf{ID-based} features refer to the item ID of user's (\textit{i.e.}, $u_i$) historical interaction sequence, \textit{i.e.}, $\bm{e}^{id}_{j}$.
    \item \textbf{Side Info-based} features contain item's attribute information and the contextual features. The former includes category, shop, brand, decay, stay time, and so on (\textit{i.e.}, $\bm{e}^{cate}_{j}, \bm{e}^{shop}_{j}, \bm{e}^{brand}_{j}, \bm{e}^{decay}_{j}, \\ \bm{e}^{stay}_{j}$, etc.), where $\bm{e}^{decay}_{j}$ represents the time interval between the user's interacted item and the current request. The latter mainly describes the interaction scene (\textit{i.e.}, $\bm{e}^{scene}_{j}$). 
\end{itemize} 
Thus, given samples consisting of users' ID-based features, Side Info-based features, and several true labels,
our model aims to predict the probabilities of \textbf{CTR} and \textbf{CVR} for a user $u_i$ with respect to the target (candidate) items $\mathcal{T}_i=\{v_j|v_j\in\mathcal{V}\}_{j=1}^{|\mathcal{T}_i|}$ as follows:
\begin{equation}
\begin{split}
\hat{y}^{ctr},\hat{y}^{cvr}=f\big(u_i,\mathcal{H}_i,\mathcal{T}_i;\Theta\big),
\end{split}
\label{}
\end{equation}
where $\hat{y}^{ctr},\hat{y}^{cvr}$ are predicted probabilities, $\Theta$ are trainable parameters, $f$ is a Transformer-based~\cite{onetrans,hyformer,mixformer} model.

Ultimately, the Transformer-based model is optimized by minimizing the standard negative log-likelihood function as follows:
\begin{equation}
\begin{split}
\mathcal{L}_{XTRs}=-\sum^{ctr,cvr}_{xtr}[y^{xtr}\mathrm{log}(\hat{y}^{xtr})+(1-y^{xtr})\mathrm{log}(1-\hat{y}^{xtr})],
\end{split}
\label{loss_xtrs}
\end{equation}
where $y^{xtr}$ denotes user's true labels, and $\hat{y}^{xtr}$ represents the predicted probabilities of XTRs.

\begin{figure*}[t!]
\begin{center}
\includegraphics[width=17.5cm]{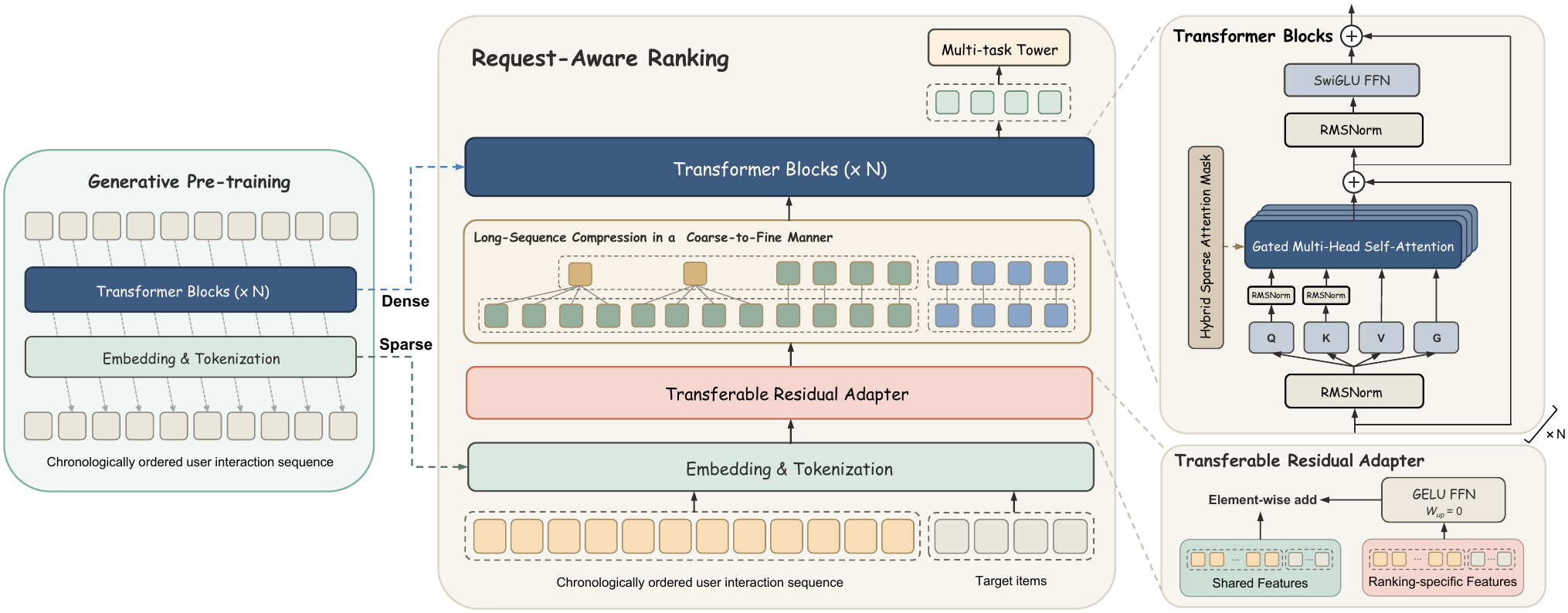}
\caption{The overall framework of LazFormer.}
\label{fig:main_model}
\end{center}
\end{figure*}

\section{Methodology}

This section introduces our proposed LazFormer, which implements a scaling Transformer framework via transferable generative pre-training, as described in Figure~\ref{fig:main_model}. Specifically, LazFormer mainly consists of three parts: 1) the \textbf{generative pre-training} aims to generate sequence features autoregressively, thus providing favorable initialization of sparse and dense parameters for the subsequent ranking process, 2) the \textbf{transferable residual adapter} draws inspiration from LoRA~\cite{lora} to inject additional ranking-specific features into the ranking process in a residual manner, thereby solving the negative transfer of
dense parameters,
and 3) the \textbf{request-aware ranking} applies a combination of long-sequence compression in a coarse-to-fine manner, hybrid sparse attention, progressive token pruning, and a request-aware paradigm to achieve model scaling alongside pre-training. Besides, we further propose an \textbf{asymmetric multi-epoch training strategy}, which continuously accumulates dense parameters across epochs for data scaling, and resets sparse parameters to alleviate overfitting.
Details of each component are demonstrated in the following parts.

\subsection{Generative Pre-training}
We utilize a Transformer architecture to autoregressively generate sequence features during the generative pre-training, providing favorable initialization of sparse and dense parameters for subsequent ranking. Unlike previous methods~\cite{gpsd,sort} that freeze pre-trained parameters, our model enables further updates and optimizations of the pre-trained parameters during ranking.

\subsubsection{\textbf{Tokenization}}
Given the historical interaction sequence $\mathcal{H}_i=\{v_j\}_{j=1}^{|\mathcal{H}_i|}$ of user $u_i$, we first transform each interacted item $v_j$ and its associated ID-based and Side Info-based features into a unified token representation. 
Specifically, for each item $v_j$, we retrieve the corresponding feature embeddings by performing embedding lookup, which can be formulated as follows:
\begin{equation}
\bm{x}_{j}
=[
\bm{e}^{id}_{j}
\Vert \bm{e}^{cate}_{j}
\Vert \bm{e}^{shop}_{j}
\Vert \bm{e}^{brand}_{j}
\Vert \bm{e}^{decay}_{j}
\Vert \bm{e}^{stay}_{j}],
\label{eq:pretrain_item_token}
\end{equation}
where $\Vert$ denotes the concatenation operation. 

Since different feature groups have distinct representation spaces, we employ linear projection and normalization to map them into the same dimension.
The final representation of the $j$-th interaction token is formulated as follows:
\begin{equation}
\bm{h}_{j}
=
\text{RMSNorm}
(
\bm{W}_{attr}\bm{x}_{j}
+
\bm{W}_{cont}\bm{e}^{scene}_{j}
),
\label{eq:pretrain_token_projection}
\end{equation}
where $\bm{W}_{attr}$ and $\bm{W}_{cont}$ are trainable parameters. Accordingly, the tokenized interaction sequence of user $u_i$ is denoted as:
\begin{equation}
\bm{H}_i
=
[
\bm{h}_{1},
\bm{h}_{2},
\ldots,
\bm{h}_{|\mathcal{H}_i|}
].
\label{eq:pretrain_token_sequence}
\end{equation}

Moreover, the sparse embedding and projection layers are shared with the subsequent ranking whenever the corresponding features are available. 


\subsubsection{\textbf{Next-Item Prediction}}

After obtaining the tokenized interaction sequence $\bm{H}_i$, we employ a stack of Transformer blocks to capture the sequential dependency among historical interactions. Specifically, each Transformer block comprises causal self-attention, a feed-forward network (FFN), normalization, and residual connections. 
Within the attention layer, we introduce pre-Norm, QKNorm, and attention gate~\cite{qwen3} to stabilize the training process.
Given the input representation $\bm{H}^{l-1}_i$ of the $l$-th layer, we first perform RMS normalization and project the normalized representations into Q/K/V matrices as follows:
\begin{equation}
\begin{split}
\bm{Q}^{l}_r, \bm{K}^{l}_r, \bm{V}^{l}_r
&=
\text{RMSNorm}
(\bm{H}^{l-1}_i\bm{W}^{Q}_{r},~ \bm{H}^{l-1}_i\bm{W}^{K}_{r},~ \bm{H}^{l-1}_i\bm{W}^{V}_{r}),\\
\end{split}
\label{eq:pretrain_qkv}
\end{equation}
where $r$ denotes the attention head, $\bm{W}^{Q}_{r}$, $\bm{W}^{K}_{r}$ and $\bm{W}^{V}_{r}$ are trainable parameters. To prevent information leakage from future interactions, we adopt a causal attention mask $\bm{M}$. Thus, the output of each attention head is calculated as follows:
\begin{equation}
\text{Head}^{l}_r
=
\text{Softmax}
\left(
\frac{\bm{Q}^l_r{\bm{K}^l_r}^{\top}}
{\sqrt{d_k}}
+
\bm{M}
\right)
\bm{V}^l_r,
\label{eq:pretrain_causal_attention}
\end{equation}
where $d_k$ is the dimension of each attention head. The outputs of all $n$ heads are concatenated and projected to obtain the multi-head attention output as follows:
\begin{equation}
\text{MHA}^l
=
[
\text{Head}^{l}_1\Vert \text{Head}^{l}_2\Vert \dots\Vert\text{Head}^{l}_n
]\bm{W}^{O},
\label{eq:pretrain_mha}
\end{equation}
where $\bm{W}^{O}$ denotes the trainable parameter.

We then apply residual connections to the attention output, followed by another normalization and a gated feed-forward network, which can be formulated as follows:
\begin{equation}
\begin{split}
\tilde{\bm{H}}^l_i
&=
\bm{H}^{l-1}_i+\text{MHA}^l,\\
\bm{H}^l_i
&=
\tilde{\bm{H}}^l_i+
\text{FFN}^l
(
\text{RMSNorm}
(\tilde{\bm{H}}^l_i)
),
\end{split}
\label{eq:pretrain_transformer_block}
\end{equation}

Ultimately, we obtain the final sequential representations
$\bm{Z}_i=[\bm{z}_1,\bm{z}_2,\dots,\bm{z}_{|\mathcal{H}_i|}]$.

To perform autoregressive generation, the hidden representation $\bm{z}_j$ is projected into the item ID embedding space to obtain $\bm{q}_j$, thus predicting the next interacted item $v_{j+1}$ as follows:
\begin{equation}
\bm{q}_j
=
\bm{W}_{auto}\bm{z}_j,
\label{eq:next_item_projection}
\end{equation}
where $\bm{W}_{auto}$ is a trainable parameter. We regard the item ID embedding $\bm{e}^{id}_{j+1}$ as the positive target and sample a set of negative items $\mathcal{N}_{i,j}$ from other interaction sequences within the same mini-batch. The probability of generating the next item is formulated as:
\begin{equation}
p(v_{j+1}\mid\mathcal{H}_{i,\leq j})
=
\frac{
\exp(s(\bm{q}_j,\bm{e}^{id}_{j+1}))
}{
\sum_{v\in\{v_{j+1}\}\cup\mathcal{N}_{i,j}}
\exp(s(\bm{q}_j,\bm{e}^{id}_{v}))
},
\label{eq:next_item_probability}
\end{equation}
where $s(\bm{a},\bm{b})=\text{sim}(\bm{a},\bm{b})/\tau$ denotes the temperature-scaled cosine similarity, and $\tau$ is a learnable temperature parameter. Accordingly, the loss function of generative pre-training is defined as follows:
\begin{equation}
\mathcal{L}_{pre}
=
-\sum_{u_i\in\mathcal{U}}
\sum_{j=1}^{|\mathcal{H}_i|-1}
\log p(v_{j+1}\mid\mathcal{H}_{i,\leq j}).
\label{eq:pretrain_loss}
\end{equation}

Through generative pre-training, LazFormer jointly learns transferable sparse item representations and dense Transformer parameters, which are subsequently used to initialize the ranking model.

\subsection{Transferable Residual Adapter}

The ranking process usually includes additional ranking-specific features that are unavailable during pre-training. Directly concatenating these features with the pre-trained representations may disturb the knowledge learned from sequential generation, leading to the negative transfer of dense parameters. 
To address this issue, we introduce a transferable residual adapter that injects additional ranking-specific features into the ranking process in a residual manner, thus transferring knowledge across pre-training and ranking.

Specifically, the additional ranking-specific features of item $v_j$ during the ranking process include add to cart, order, and gap (\textit{i.e.}, $\bm{e}^{atc}_{j}, \bm{e}^{order}_{j}, \bm{e}^{atc-gap}_{j}, \bm{e}^{order-gap}_{j}$), where gap denotes the time interval between adding to cart or ordering and the most recent click. These features provide an efficient way to incorporate multi-behavior signals beyond clicks without explicitly extending the interaction sequence, thereby avoiding the substantial computational overhead brought by longer self-attention inputs.
The transferable residual adapter can be formulated as:
\begin{equation}
\begin{split}
\bm{s}_j &= [\bm{e}^{atc}_{j} \Vert \bm{e}^{order}_{j} \Vert \bm{e}^{atc-gap}_{j}\Vert \bm{e}^{order-gap}_{j}],\\
\bm{a}_j
&=
\bm{W}_{up}
\text{GELU}
(
\bm{W}_{down}\bm{s}_j
),
\end{split}
\label{eq:residual_adapter}
\end{equation}
where $\bm{a}_j$ denotes the residual representation, $\bm{W}_{down}$ and $\bm{W}_{up}$ are trainable parameters. Subsequently, $\bm{a}_j$ is added to the interaction token loaded from the generative pre-training as follows:
\begin{equation}
\bm{h}^{rank}_j
=
\text{RMSNorm}
(
\bm{W}_{attr}\bm{x}_j
+
\bm{W}_{cont}\bm{e}^{scene}_j
+
\bm{a}_j
),
\label{eq:ranking_adapter_token}
\end{equation}
where $\bm{h}^{rank}_j$ represents the interaction token for the subsequent ranking process.

To ensure training stability during the initial phase of the ranking process,
we initialize $\bm{W}_{up}=\bm{0}$. Therefore, the adapter initially produces a zero residual as follows:
\begin{equation}
\bm{a}_j=\bm{0}
~\to~
\bm{h}^{rank}_j
=
\text{RMSNorm}
(
\bm{W}_{attr}\bm{x}_j
+
\bm{W}_{cont}\bm{e}^{scene}_j
),
\label{eq:adapter_initialization}
\end{equation}
This zero initialization makes the ranking process start from the features learned during generative pre-training, while gradually incorporating ranking-specific features during optimization. Unlike methods~\cite{sort,gpsd} that freeze the pre-trained parameters, LazFormer utilizes the transferable residual adapter to enable the transfer of sparse and dense parameters, allowing the pre-trained knowledge to adapt to the ranking objective.

\subsection{Request-Aware Ranking}

After generative pre-training, we utilize a transferable residual adapter to transfer the learned sparse and dense parameters to initialize the ranking process. However, the ranking process typically involves substantially longer interaction sequences and multiple target items within the same request. 
If we continue to train on individual target items, the same long sequence will be encoded repeatedly. As the sequence length increases, storage and bandwidth become bottlenecks.
Thus, to efficiently scale Transformers to such an industrial ranking process, we introduce request-aware ranking, which consists of long-sequence compression in a coarse-to-fine manner, hybrid sparse attention, and a request-aware paradigm.

\subsubsection{\textbf{Long-Sequence Compression in a Coarse-to-Fine Manner}}

Users' interaction sequence usually contains both fine-grained short-term preference and relatively coarse-grained long-term preference. Directly encoding all historical interactions results in excessive computational and memory costs. Consequently, we preserve the short-term interactions while compressing the remaining long-term interactions through group-wise sum pooling.

Let $L_i=|\mathcal{H}_i|$ denote the original historical interaction sequence length and $\ell_i=\min(L_i,L_s)$ denote the number of short-term tokens retained, where $L_s$ is a predefined short-term sequence length. We divide the remaining interactions into groups of size $g$, and calculate the representation of the $k$-th group as follows:
\begin{equation}
\bm{c}_{k}
=
\sum_{j=\ell_i+(k-1)g+1}^{\min(\ell_i+kg,L_i)}
\bm{h}^{rank}_j,
\label{eq:long_sequence_pooling}
\end{equation}
The compressed historical sequence can be formulated as:
\begin{equation}
\begin{split}
\hat{\bm{H}}^{rank}_i
=
[
\bm{c}_{1},\ldots,\bm{c}_{N_i},\bm{h}^{rank}_1,\ldots,\bm{h}^{rank}_{\ell_i}
],
\end{split}
\label{eq:compressed_history}
\end{equation}
where $N_i=\left\lceil\frac{L_i-\ell_i}{g}\right\rceil$ denotes the number of compressed long-term groups. Consequently, the historical interaction sequence length of $u_i$ is reduced from $L_i$ to $N_i+\ell_i$, while the fine-grained short-term preference are retained.

\begin{figure}[t!]
\begin{center}
\includegraphics[width=7.5cm]{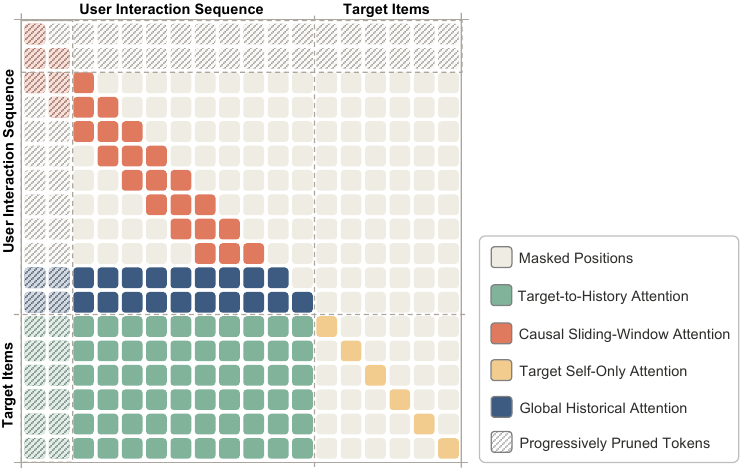}
\caption{An illustration of hybrid sparse attention mask.}
\label{fig:hybrid_sparse_attention}
\end{center}
\end{figure}

\subsubsection{\textbf{Hybrid Sparse Attention}}

Based on the compressed historical sequence, we append all target item tokens belonging to the same request to construct the ranking input as follows:
\begin{equation}
\begin{split}
\bm{X}^{0}_i
&=
[
\hat{\bm{H}}^{rank}_i
\Vert
\bm{T}_i
],\\
&=[
\bm{c}_{1},\ldots,\bm{c}_{N_i},\bm{h}^{rank}_1,\ldots,\bm{h}^{rank}_{\ell_i}, \bm{t}_{1},\ldots,\bm{t}_{|\mathcal{T}_i|}
],
\label{eq:ranking_input_sequence}
\end{split}
\end{equation}
where $\hat{\bm{H}}^{rank}_i$ denotes the compressed historical sequence, and $\bm{T}_i$ represents the target item tokens, which are constructed using the shared feature embeddings and projection layers.

To facilitate long-range information propagation, we designate the \emph{last} $G_i$ history
tokens, i.e., the $G_i$ most recent behaviors, as global tokens, and treat the remaining tokens as local historical tokens.
Thus, the global historical tokens retain user's short-term fine-grained preference.

Since the full self-attention has quadratic cost,
we construct a hybrid sparse attention (as shown in Figure~\ref{fig:hybrid_sparse_attention}) pattern by combining causal sliding-window attention and global historical tokens. Specifically, let $p$ and $q$ denote the query and key positions, the hybrid sparse attention relation can be defined as follows:
\begin{equation}
\mathcal{A}_i(p,q)=
\begin{cases}
1, & p,q\leq \tilde{L}_i,~q\leq p,
~(p-q<W ~\text{or}~ p>\tilde{L}_i-G_i),\\
1, & p>\tilde{L}_i,~q\leq \tilde{L}_i,\\
1, & p>\tilde{L}_i,~q>\tilde{L}_i,~p=q,\\
0, & \text{otherwise},
\end{cases}
\label{eq:sparse_attention_relation}
\end{equation}
where $W$ denotes the sliding-window size, $\tilde{L}_i=N_i+\ell_i$ denotes the number of history tokens after compression. 
Specifically, each local token attends to its preceding local window, while global tokens enable long-range propagation. The global tokens provide persistent information anchors for information propagation across distant historical interactions. Each target item attends to all valid historical tokens, including both global and local ones. In contrast, a target item can attend only to itself within the target sequence, preventing information leakage across candidate predictions.

Consequently, the corresponding attention mask $\bm{M}_{p,q}$ is constructed as follows:
\begin{equation}
\bm{M}_{p,q}
=
\begin{cases}
0, & \mathcal{A}_i(p,q)=1,\\
-\infty, & \mathcal{A}_i(p,q)=0,
\end{cases}
\label{eq:sparse_attention_mask}
\end{equation}
The hybrid sparse attention operation is then formulated as:
\begin{equation}
\operatorname{SparseAttn}(\bm{Q},\bm{K},\bm{V})
=
\operatorname{Softmax}
\left(
\frac{\bm{Q}\bm{K}^{\top}}{\sqrt{d_k}}
+
\bm{M}
\right)\bm{V},
\label{eq:ranking_sparse_attention}
\end{equation}
Compared with full self-attention, the hybrid sparse attention pattern avoids unnecessary historical interaction pairs while retaining local sequential dependencies, global information propagation, and complete target to history interactions. We materialise this sparsity via flex attention~\cite{flexatt}, which compiles the block-sparse pattern into fused CUDA kernels. 

Moreover, based on the compressed historical sequence, we progressively prune historical tokens across Transformer layers to reduce computation in deeper layers. Let $N_i^l$ denote the number of historical tokens retained at the $l$-th layer. We linearly decrease the historical sequence length from $\tilde{L}_i$ to a predefined minimum length $N_{\min}$ as follows:
\begin{equation}
N_i^l
=
\min(
\tilde{L}_i,
\max(
N_{\min},
\operatorname{Round}
[
\tilde{L}_i-\frac{l-1}{L-1}(\tilde{L}_i-N_{\min})
]
)
),
\label{eq:token_pruning_schedule}
\end{equation}
where $L$ denotes the number of Transformer layers. After the $l$-th layer, only the \emph{most recent} $N_i^{l+1}$ history representations and all target-item representations are propagated to the next layer. 
Consequently, the combination of long-sequence compression, hybrid sparse attention, and progressive token pruning enables LazFormer to reduce the computation required by long-sequence modeling. 

\subsubsection{\textbf{Request-Aware Paradigm}}

Industrial logs contain multiple target items within the same request. Traditional ranking models construct an independent sample $(u_i,\mathcal{H}_i,v_j,y_{ij})$ for each target item $v_j$. Consequently, the same historical interaction sequence $\mathcal{H}_i$ is repeatedly serialized, stored, transferred from CPU to GPU, and encoded for all target items within the request. As the length of $\mathcal{H}_i$ increases, redundant data movement and user-side computation become substantial system bottlenecks, leading to a waste of computational resources.

To address this issue, we organize the training samples at the request level, which can be formulated as follows:
\begin{equation}
\mathcal{R}_i
=
\left(
u_i,
\mathcal{H}_i,
\mathcal{T}_i,
\mathcal{Y}_i
\right),
\label{eq:request_level_sample}
\end{equation}
where all target items $\mathcal{T}_i$ and their labels $\mathcal{Y}_i$ are organized within the same request, sharing a single historical sequence $\mathcal{H}_i$. Instead of independently encoding $\mathcal{H}_i$ for each target item, LazFormer serializes and transfers $\mathcal{H}_i$ only once and simultaneously processes all target items. 

This request-aware paradigm serves as a system-side complement to hybrid sparse attention: hybrid sparse attention reduces the computational cost of long-sequence modeling, while request-level training removes redundant storage, data transfer, and repeated history encoding.

After passing the input $\bm{X}^{0}_i$ through the ranking Transformer, we extract the contextualized representations of all target items as:
\begin{equation}
\bm{Z}^{t}_i
=
[
\bm{z}^{t}_{1},
\bm{z}^{t}_{2},
\ldots,
\bm{z}^{t}_{|\mathcal{T}_i|}
].
\label{eq:request_target_outputs}
\end{equation}
Each target representation is further fed into the corresponding prediction network as follows:
\begin{equation}
\hat{y}^{xtr}=f_{xtr}(\bm{Z}^{t}_i),
\label{eq:request_aware_prediction}
\end{equation}
where $f_{xtr}$ denotes the task-specific prediction network. Since target items are isolated by the hybrid sparse attention mask, the request-aware paradigm preserves the independence of individual target predictions while sharing the storage, transfer, and encoding of the same user history.

Ultimately, the ranking process is optimized by the XTRs prediction loss defined in Eq.~(\ref{loss_xtrs}) as follows:
\begin{equation}
\mathcal{L}_{rank}
=
\mathcal{L}_{XTRs},
\label{eq:ranking_loss}
\end{equation}
During the ranking process, the sparse and dense parameters are jointly optimized with the transferable residual adapter and prediction networks.

\subsection{Asymmetric Multi-Epoch Training Strategy}

Although generative pre-training provides favorable initialization for both sparse and dense parameters, a single epoch for ranking model may still be insufficient for the dense backbone to fully absorb the supervision contained in large-scale interaction logs. A straightforward solution is to train the ranking model for multiple epochs. However, prior studies~\cite{meda,meda_v2} have observed a one-epoch overfitting phenomenon, 
indicating that repeatedly updating sparse embeddings on the same training data may cause overfitting. To address this issue, we propose an asymmetric multi-epoch training strategy: sparse parameters are re-initialized to their pre-trained state at the beginning of each epoch, while dense parameters inherit the optimized state from the previous epoch.

Specifically, we divide the parameters of LazFormer into sparse parameters $\Theta^{s}$ and dense parameters $\Theta^{d}$. 
At the beginning of each epoch, we reset $\Theta^{s}$ to its initial state $\Theta^{s}_{init}$, which is transferred from generative pre-training. In contrast, the dense parameters inherit the optimized state from the previous epoch, which can be formulated as follows:
\begin{equation}
\begin{split}
\Theta^{s}_{e,0}
&=
\Theta^{s}_{init},\\
\Theta^{d}_{e,0}
&=
\begin{cases}
\Theta^{d}_{init}, & e=1,\\
\Theta^{d}_{e-1,K_{e-1}}, & e>1,
\end{cases}
\end{split}
\label{eq:multi_epoch_initialization}
\end{equation}
where $e$ denotes the epoch index, and $K_{e-1}$ denotes the number of training steps in epoch $e-1$. During each epoch, both sparse and dense parameters are jointly optimized by $\mathcal{L}_{rank}$.

Consequently, dense parameters can continuously absorb ranking supervision across epochs, while resetting sparse parameters helps preserve the transferable sparse representations learned during pre-training and alleviate overfitting. In this way, the proposed asymmetric multi-epoch training strategy improves effective data scaling while maintaining transferable sparse knowledge.



\section{Experiments}

In this section, we conduct extensive offline experiments and online A/B testing on our industrial recommendation system to answer the following research questions:
\begin{itemize}
    \item \textbf{RQ1:} Can LazFormer outperform existing scalable ranking Transformers and pre-training-ranking paradigms?
    \item \textbf{RQ2:} Does the transferable residual adapter effectively bridge generative pre-training and downstream ranking?
    \item \textbf{RQ3:} Does the asymmetric multi-epoch training strategy enable effective data scaling while avoiding sparse overfitting?
    \item \textbf{RQ4:} Can LazFormer continuously benefit from scaling model capacity, sequence length, and training data under practical efficiency constraints?
    \item \textbf{RQ5:} How do the efficiency-oriented designs of request-aware ranking affect the trade-off between recommendation performance and training efficiency?
    \item \textbf{RQ6:} How does LazFormer perform in real-world online A/B testing under industrial deployment settings?
\end{itemize}

\subsection{Experimental Settings}
\subsubsection{\textbf{Datasets}}

For offline evaluation, we use large-scale industrial data from a real-world e-commerce recommendation platform. The data consist of a generative pre-training dataset and a ranking dataset. Following~\cite{gpsd}, the pre-training dataset is constructed by chronologically sorting each user's interactions over one year and segmenting them into subsequences. The ranking dataset is built from chronologically ordered user interaction sequences, using logs from the past 25 consecutive days for training and the subsequent day for testing. Table~\ref{tab:datasets} summarizes the dataset statistics.




\begin{table}[t]
\centering
\caption{Statistics of datasets.}
\label{tab:datasets}
\setlength{\tabcolsep}{5pt}
\begin{tabular*}{0.47\textwidth}
{@{\extracolsep{\fill}}@{}lccc@{}}
\toprule
\textbf{Dataset} & \textbf{$\#$~Users} & \textbf{$\#$~Tokens} & \textbf{$\#$~Impressions} \\
\midrule
Pre-training & 16M & 11B & -- \\
Ranking & 11M & -- & 370M \\
\bottomrule
\end{tabular*}
\end{table}

\subsubsection{\textbf{Baselines}}
We compare \textbf{LazFormer} with representative baselines covering vanilla Transformer (\textit{i.e.}, \textbf{Transformer}), scalable ranking Transformers (\textit{i.e.}, \textbf{HSTU}~\cite{hstu}, \textbf{OneTrans}~\cite{onetrans}), pre-training and ranking paradigm (\textit{i.e.}, \textbf{SORT}~\cite{sort}), and efficient long-sequence modeling (\textit{i.e.}, \textbf{STCA}~\cite{stca}). For fair comparison, all baselines are initialized with the same pre-trained sparse parameters in our experiments, although the original ranking-only baselines are typically trained from scratch in prior work. Only LazFormer further transfers dense parameters from generative pre-training.

\subsubsection{\textbf{Evaluation Metrics}}

Following previous works~\cite{onetrans,mixformer,farm,dmf}, we evaluate the effectiveness of LazFormer and baselines using two widely adopted metrics, including AUC and GAUC.


For efficiency and scalability evaluation, we further report the number of trainable model parameters and floating-point operations (\textit{i.e.}, FLOPs),
which serve as proxies for computational complexity and indicate the scalability and deployment cost of different models in real-world industrial recommendation systems. 

\subsubsection{\textbf{Implementation Details}}

We implement LazFormer with PyTorch on NVIDIA H100 GPUs using AdamW~\cite{adam}. Unless otherwise specified, Transformer, HSTU, OneTrans, SORT, STCA, and LazFormer$^*$ share the same backbone configuration: 3 Transformer layers, hidden size 256, 4 attention heads, FFN dimension 1,024, and sequence length 1,024. SORT uses one shared expert and four routed experts with two activated experts.

Generative pre-training uses a 5-layer Transformer with item-ID embedding dimension 64 and temperature coefficient $\tau=0.07$ with 16,000 in-batch negatives. Compared with LazFormer$^*$, LazFormer scales the ranking architecture to 5 Transformer layers and sequence length 2,048, where the number of retained short-term tokens is $\ell_i=1{,}024$,
the compression window size is $g=8$, and $W$, $G_i$, and $N_{\min}$ are all set to 128. All models are trained for one epoch by default, while LazFormer$^\dagger$ denotes LazFormer trained with the asymmetric multi-epoch training strategy for three epochs (please refer to Section~\ref{rq3}).

\begin{table}[t]
\centering
\begin{threeparttable}
\caption{Offline performance in terms of AUC and GAUC, where the best results are in bold and the best baseline ones are underlined. Indented rows are derived from LazFormer. All baselines load the same pre-trained sparse parameters, while LazFormer further transfers the dense ones. Results are averaged over three runs.}
\label{exp_offline}
\setlength\tabcolsep{2.0pt}
\begin{tabular*}{\columnwidth}{@{\extracolsep{\fill}}@{}lcccccc@{}}
\toprule
\multirow{2.5}{*}{\textbf{Models}\tnote{a}} & \multicolumn{2}{c}{\textbf{CTR}}  & \multicolumn{2}{c}{\textbf{CVR}} & \multicolumn{2}{c}{\textbf{Efficiency}\tnote{b}}
\\ \cmidrule(r){2-3} \cmidrule(r){4-5} \cmidrule(r){6-7} & \textbf{AUC} & \textbf{GAUC} & \textbf{AUC} & \textbf{GAUC} & \textbf{Params} & \textbf{FLOPs} \\
\midrule
Transformer & 0.7630 & 0.6704 & 0.8835 & 0.7001 & 6M & 20G \\
HSTU & 0.7636 & 0.6725 & 0.8826 & 0.6942 & 6M & 22G \\
STCA & 0.7636 & 0.6732 & 0.8830 & 0.6996 & 7M & 3G \\
OneTrans & \underline{0.7661} & 0.6738 & 0.8840 & 0.7053 & 28M & 15G \\
SORT & 0.7652 & \underline{0.6740} & \underline{0.8850} & \underline{0.7068} & 16M & 29G \\
\midrule
\textbf{LazFormer} & 0.7735 & 0.6868 & 0.8905 & 0.7130 & 9M & 25G \\
\quad\textit{Improve} (pt) $\uparrow$ & +0.74 & +1.28 & +0.55 & +0.62 & -- & -- \\
\quad LazFormer$^*$ & 0.7704 & 0.6808 & 0.8885 & 0.7094 & 7M & 14G \\
\quad LazFormer$^\dagger$ & \textbf{0.7761} & \textbf{0.6891} & \textbf{0.8933} & \textbf{0.7143} & 9M & 75G \\
\bottomrule
\end{tabular*}
\begin{tablenotes}[flushleft]
\footnotesize
\item[a] LazFormer: our default configuration (5 layers, sequence length 2,048) used in all subsequent experiments. LazFormer$^*$: downscaled to the baseline configuration (3 layers, sequence length 1,024). LazFormer$^\dagger$: with asymmetric multi-epoch training. \textit{Improve}: absolute gain of \mbox{LazFormer} over the underlined best baseline of each metric in percentage points.
\item[b] Params exclude embedding parameters, amounting to about 5B for all models. FLOPs are measured for one training pass (forward and backward) over same request-level sample and are accumulated over epochs. FLOPs count unmasked attention pairs only.
\end{tablenotes}
\end{threeparttable}
\vspace{-0.3cm}
\end{table}

\subsection{Offline Performance (RQ1)}

Table~\ref{exp_offline} reports the offline performance of LazFormer and representative baselines. Note that for our industrial dataset, an improvement of 0.1\,pt (\textit{i.e.}, 0.001) in AUC or GAUC is already considered significant for CTR prediction, as also observed in~\cite{din,twin}. LazFormer$^*$ already outperforms all baselines by a clear margin, even under the aligned depth and sequence length. It also requires fewer training FLOPs than the vanilla Transformer because of the hybrid sparse attention and progressive token pruning. The full LazFormer$^\dagger$ further achieves the best overall performance.

First, compared with scalable ranking Transformers (\textit{i.e.}, HSTU and OneTrans), the gains of LazFormer are not merely from larger computation budgets, as evidenced by LazFormer$^*$ already outperforming them under the aligned configuration. Moreover, since all baselines are also initialized with the same pre-trained sparse parameters, the results show that the advantage of LazFormer does not come from sparse-only pre-trained initialization alone, but from going beyond sparse-only transfer toward transferable dense adaptation.
Second, even compared with SORT, which also leverages pre-trained knowledge via sparse-only transfer, LazFormer's transferable dense adaptation yields further gains.
Finally, compared with the efficient long-sequence baseline STCA, LazFormer consistently achieves better performance. This result demonstrates the effectiveness of the proposed coarse-to-fine long-sequence modeling for capturing both short-term and long-term user preferences. Although STCA requires the fewest FLOPs by aggressively compressing the user history, it sacrifices fine-grained short-term preferences and thus underperforms LazFormer$^*$, which we further analyze in Section~\ref{rq5}.

\begin{table}[t]
\centering
\caption{Ablation study of the transferable residual adapter. Sparse denotes sparse-only transfer.
NoFeat removes ranking-specific features and keeps only shared features.
Direct directly fuses ranking-specific features with shared features.
PreProj uses a pre-projection fusion layer before the transferred input projection.}
\label{exp_adapter}
\setlength\tabcolsep{2pt}{
\begin{tabular*}{0.47 \textwidth}{@{\extracolsep{\fill}}@{}lcccc@{}}
\toprule
\multirow{2.5}{*}{\bf Variants} &
\multicolumn{2}{c}{\textbf{CTR}} &
\multicolumn{2}{c}{\textbf{CVR}} \\
\cmidrule(r){2-3}\cmidrule(r){4-5}&
\textbf{AUC} & \textbf{GAUC} & \textbf{AUC} & \textbf{GAUC} \\
\midrule
Sparse
& 0.7699 & 0.6810 & 0.8898 & 0.7109 \\
NoFeat
& 0.7694 & 0.6843 & 0.8809 & 0.7038 \\
Direct
& 0.7701 & 0.6817 & 0.8897 & 0.7091 \\
PreProj
& 0.7699 & 0.6810 & 0.8889 & 0.7050 \\
LazFormer
& \textbf{0.7735} & \textbf{0.6868} & \textbf{0.8905} & \textbf{0.7130} \\
\bottomrule
\end{tabular*}
}
\end{table}

\subsection{Transferable Residual Adapter Analysis (RQ2)}

To investigate whether LazFormer effectively bridges generative pre-training and downstream ranking under inconsistent feature spaces, we compare four variants that differ in whether dense parameters are transferred and how shared features and ranking-specific features are fused.

\begin{itemize}
    \item \textbf{Sparse:} transfers only the pre-trained sparse parameters, while keeping the ranking model unchanged.
    \item \textbf{NoFeat:} transfers both sparse and dense parameters but removes the ranking-specific features, so the ranking input contains only the shared features used in pre-training.
    \item \textbf{Direct:} transfers both sparse and dense parameters, and directly fuses ranking-specific features with shared features at the tokenization stage.
    \item \textbf{PreProj:} transfers both sparse and dense parameters, and uses a pre-projection layer to fuse ranking-specific features with shared features before the transferred input projection.
\end{itemize}

As shown in Table~\ref{exp_adapter}, LazFormer consistently achieves the best performance. Compared with \textbf{Sparse}, the gain demonstrates the benefit of transferring dense parameters beyond sparse embeddings. Interestingly, \textbf{NoFeat} achieves a relatively strong CTR GAUC, suggesting that dense parameter transfer remains effective when ranking is performed using only shared features aligned with pre-training. However, its CVR performance drops substantially, indicating that shared features alone are insufficient for conversion modeling. This result suggests that ranking-specific features, especially those related to stronger purchase-intent signals such as $\bm{e}^{order}_{j}$ and $\bm{e}^{order-gap}_{j}$, provide important task-specific cues for downstream conversion prediction and should not be removed merely to better align the ranking input with pre-training.

Compared with \textbf{Direct} and \textbf{PreProj}, LazFormer further shows that not only the presence of ranking-specific features but also their fusion manner is crucial for stable transfer: directly fusing ranking-specific features perturbs the input representation space learned during pre-training, while an additional randomly initialized pre-projection still perturbs the input representations fed into the pre-trained dense backbone. In contrast, the transferable residual adapter preserves the pre-trained representation at initialization via the zero-initialized residual branch and gradually incorporates ranking-specific features during optimization, thus alleviating the negative transfer of dense parameters.

\begin{table}[t]
\centering
\caption{Study of the asymmetric multi-epoch training strategy, where ``freeze'', ``continue'', and ``reset'' denote sparse freezing, continuous sparse updating, and sparse reset.}
\label{tab:multi_epoch}
\setlength\tabcolsep{2pt}{
\begin{tabular*}{0.47 \textwidth}{@{\extracolsep{\fill}}@{}lccccc@{}}
\toprule
\multirow{2.5}{*}{\bf Config} &
\multicolumn{2}{c}{\textbf{Loading Params}} & \multirow{2.5}{*}{\bf Epoch} &
\multicolumn{2}{c}{\textbf{CTR}} \\
\cmidrule(r){2-3}\cmidrule(r){5-6}&
\textbf{Sparse} & \textbf{Dense} & & \textbf{AUC} & \textbf{GAUC} \\
\midrule
Config-1
& \ding{55} & \ding{55} & 1 & 0.7577 & 0.6587 \\
Config-2
& \ding{51}, freeze & \ding{55} & 1 & 0.7675 & 0.6775 \\
Config-3
& \ding{51}, continue & \ding{55} & 1 & 0.7699 & 0.6810 \\
Config-4
& \ding{51}, continue & \ding{55} & 3 & 0.7387 & 0.6694 \\
Config-5
& \ding{51}, reset & \ding{55} & 3 & 0.7731 & 0.6851 \\
Config-6
& \ding{51}, reset & \ding{51} & 3 & \textbf{0.7761} & \textbf{0.6891} \\
\bottomrule
\end{tabular*}
}
\end{table}

\subsection{Asymmetric Multi-Epoch Training Strategy Analysis (RQ3)} \label{rq3}

To investigate the effectiveness of the asymmetric multi-epoch training strategy, we compare different parameter loading and optimization strategies in Table~\ref{tab:multi_epoch}. Specifically, compared with Config-1 trained from scratch, Config-2 and Config-3 achieve substantial improvements by loading pre-trained sparse parameters. Moreover, Config-3 consistently outperforms Config-2, indicating that adapting sparse parameters to the ranking objective is more effective than freezing them during one-epoch training.
We also find that continuously optimizing sparse parameters for three epochs in Config-4 leads to severe performance degradation, and Config-5 substantially outperforms Config-4 by resetting sparse parameters to their pre-trained state at the beginning of each epoch. This suggests that repeatedly updating enormous sparse parameters on the same ranking data may cause overfitting and disturb the transferable sparse knowledge. 
Finally, Config-6 further loads the pre-trained dense parameters and achieves the best performance. This verifies that our asymmetric multi-epoch training strategy enables dense parameters to continuously absorb ranking supervision from training data while preventing the overfitting of sparse parameters. 

\begin{table}[t]
\centering
\caption{Different model scales of LazFormer.}
\label{tab:scaling_laws}
\setlength{\tabcolsep}{5pt}
\begin{tabular*}{0.47\textwidth}
{@{\extracolsep{\fill}}@{}lcccc@{}}
\toprule
\textbf{Scale} & \textbf{Depth} & \textbf{Width} & \textbf{Intermediate Size} &
\textbf{Params} \\
\midrule
Small & 5 & 256 & 1024 & 9M  \\
Medium & 8 & 384 & 1536 & 23M  \\
Large & 12 & 512 & 2048 & 55M  \\
\bottomrule
\end{tabular*}
\end{table}

\begin{figure}[t]
\setlength{\abovecaptionskip}{0.cm}
	\begin{center}
        \subfigure
        {\begin{minipage}[b]{.49\linewidth}
        \centering
        \includegraphics[scale=0.15]{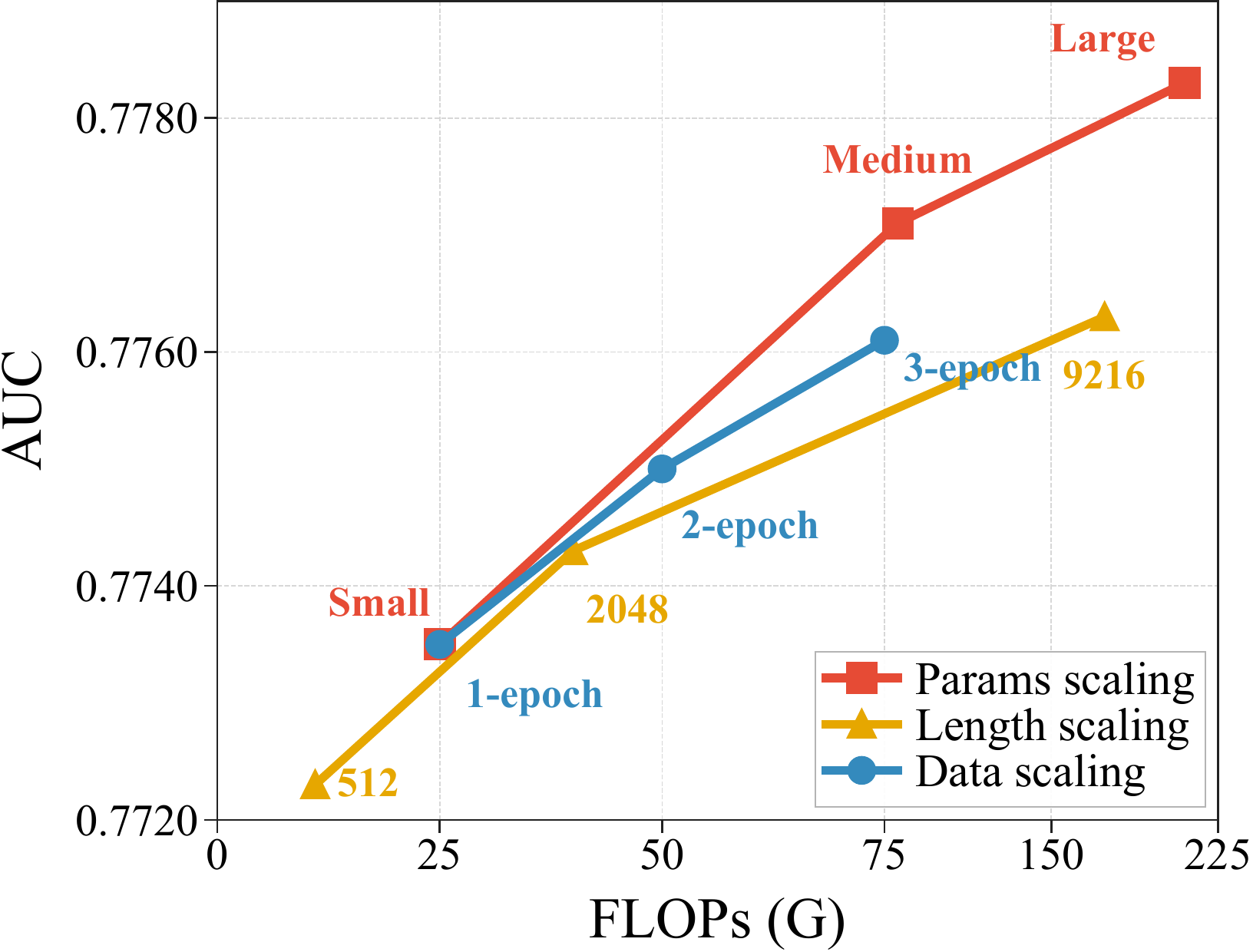}
        \end{minipage}}
        \subfigure
        {\begin{minipage}[b]{.49\linewidth}
        \centering
        \includegraphics[scale=0.15]{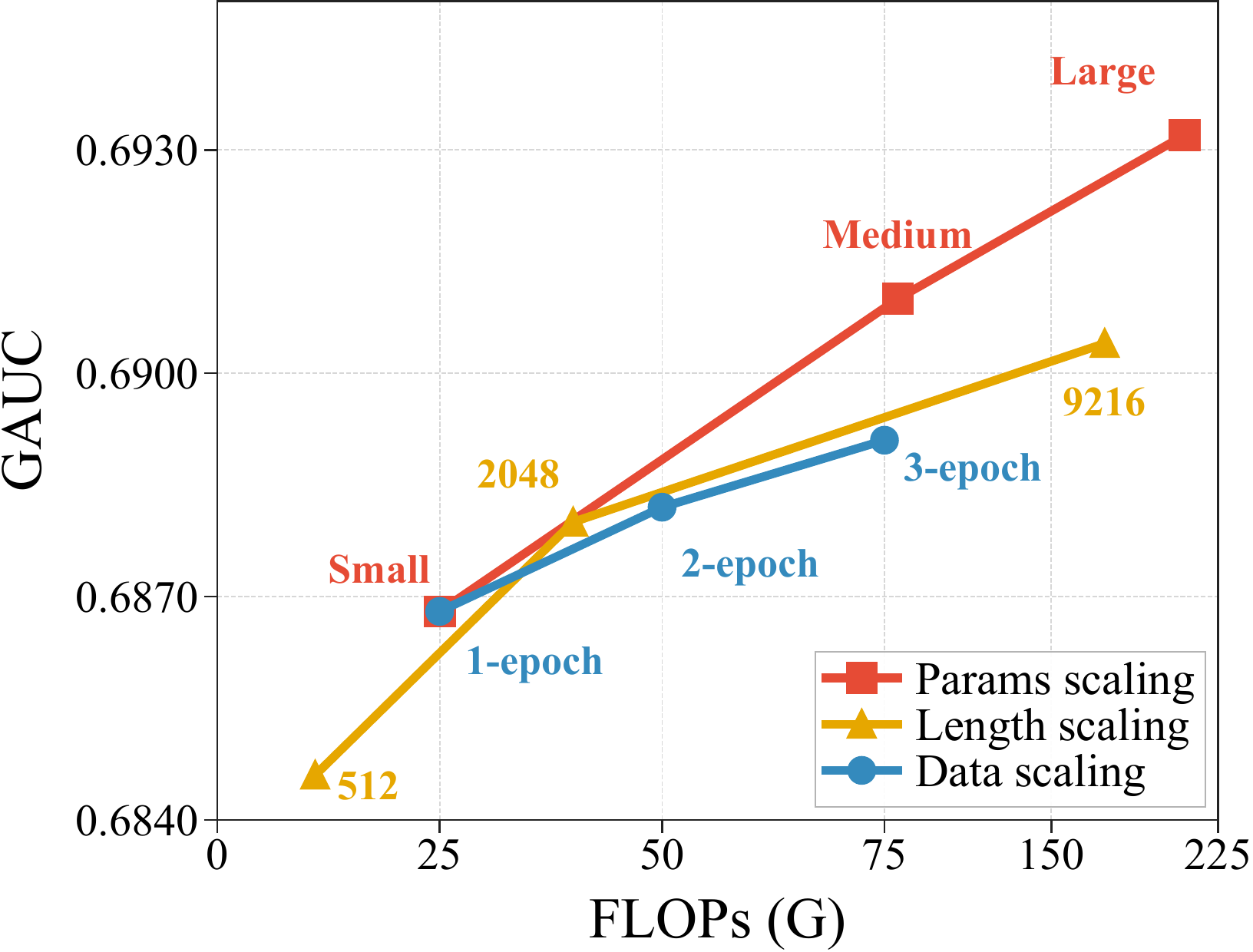}
        \end{minipage}}
	\caption{Scaling Laws of LazFormer. Performance scales monotonically with compute (FLOPs) across parameters, sequence length, and training data.}
	\label{fig:scaling_laws}
	\end{center}
\vspace{-0.2cm}
\end{figure}

\subsection{Scaling Laws Analysis (RQ4)}

To investigate the scalability of LazFormer, we conduct experiments along three dimensions: model capacity (\textit{i.e.}, parameters), training data, and sequence length. As shown in Table~\ref{tab:scaling_laws}, we scale the model by jointly increasing its depth, width, and intermediate size. For data scaling, we train LazFormer for one, two, and three epochs using the asymmetric multi-epoch training strategy. For sequence length scaling, we increase the historical sequence length from 512 to 9,216. The results are presented in Figure~\ref{fig:scaling_laws}.

We observe that both AUC and GAUC improve monotonically as the computational budget increases across all three scaling dimensions. Specifically, scaling LazFormer from Small to Medium and Large consistently improves performance, demonstrating that the Transformer backbone can effectively benefit from increased model capacity. Moreover, training for more epochs yields continuous gains. This observation validates that our asymmetric multi-epoch training strategy enables dense parameters to absorb more ranking supervision while preserving transferable sparse knowledge. Besides, increasing the historical sequence length also steadily improves performance, indicating that longer interaction sequences provide valuable signals for modeling user preferences. Meanwhile, long-sequence compression and hybrid sparse attention make sequence scaling computationally feasible. Overall, these results demonstrate the favorable scaling capability of LazFormer and verify that co-scaling model capacity and training data can continuously improve industrial recommendation performance.


\begin{table}[t]
\centering
\caption{Configurations of different long-sequence compression strategies. $\ell_i$ denotes the number of retained short-term tokens, $g$ denotes the group size for long-term compression, and $\ell_i+N_i$ denotes the sequence length after compression.}
\label{tab:compression_strategies}
\setlength{\tabcolsep}{5pt}
\begin{tabular*}{0.47\textwidth}
{@{\extracolsep{\fill}}@{}lccccccc@{}}
\toprule
\textbf{Strategy} &
\textbf{1} &
\textbf{2} &
\textbf{3} &
\textbf{4} &
\textbf{5} &
\textbf{6} &
\textbf{7} \\
\midrule
\textbf{$\ell_i$}
& 0 & 256 & 512 & 0 & 1024 & 1024 & 2048 \\
\textbf{$g$}
& 4 & 8 & 16 & 2 & 16 & 8 & -- \\
\textbf{$\ell_i+N_i$}
& 512 & 480 & 608 & 1024 & 1088 & 1152 & 2048 \\
\bottomrule
\end{tabular*}
\end{table}

\begin{figure}[t!]
\begin{center}
\includegraphics[width=8.5cm]{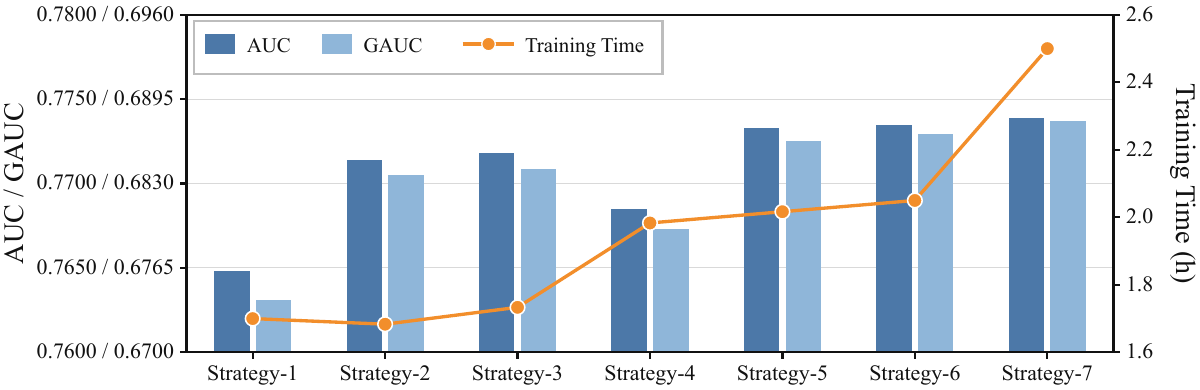}
\caption{Comparison of different long-sequence compression strategies in terms of CTR prediction and training time.}
\label{fig:compression_strategy}
\end{center}
\end{figure}


\subsection{Efficiency Analysis of Request-Aware Ranking (RQ5)}
\label{rq5}


To investigate the efficiency-performance trade-off of the efficiency-oriented designs in request-aware ranking, we analyze long-sequence compression and hybrid sparse attention in turn. First, we compare seven long-sequence compression strategies listed in Table~\ref{tab:compression_strategies}, with the results shown in Figure~\ref{fig:compression_strategy}. We make the following observations.

First, directly compressing the entire sequence leads to suboptimal performance. In particular, although Strategy-4 retains 1,024 tokens after compression, it still underperforms Strategy-2 and Strategy-3, suggesting that fine-grained short-term sequences are essential for accurately capturing user preferences.
Second, among the coarse-to-fine strategies, retaining more short-term interactions generally leads to better performance. Specifically, Strategy-6 (\textit{i.e.}, LazFormer) preserves 1,024 short-term tokens and compresses the remaining interactions with a group size of 8, achieving the best performance among all compression strategies. Compared with the uncompressed Strategy-7, it incurs only a slight performance drop while significantly reducing training time. 
Overall, Strategy-6 achieves the best trade-off between recommendation performance and training efficiency, validating the effectiveness of the proposed coarse-to-fine long-sequence compression strategy.


\begin{figure}[t!]
\begin{center}
\includegraphics[width=8.5cm]{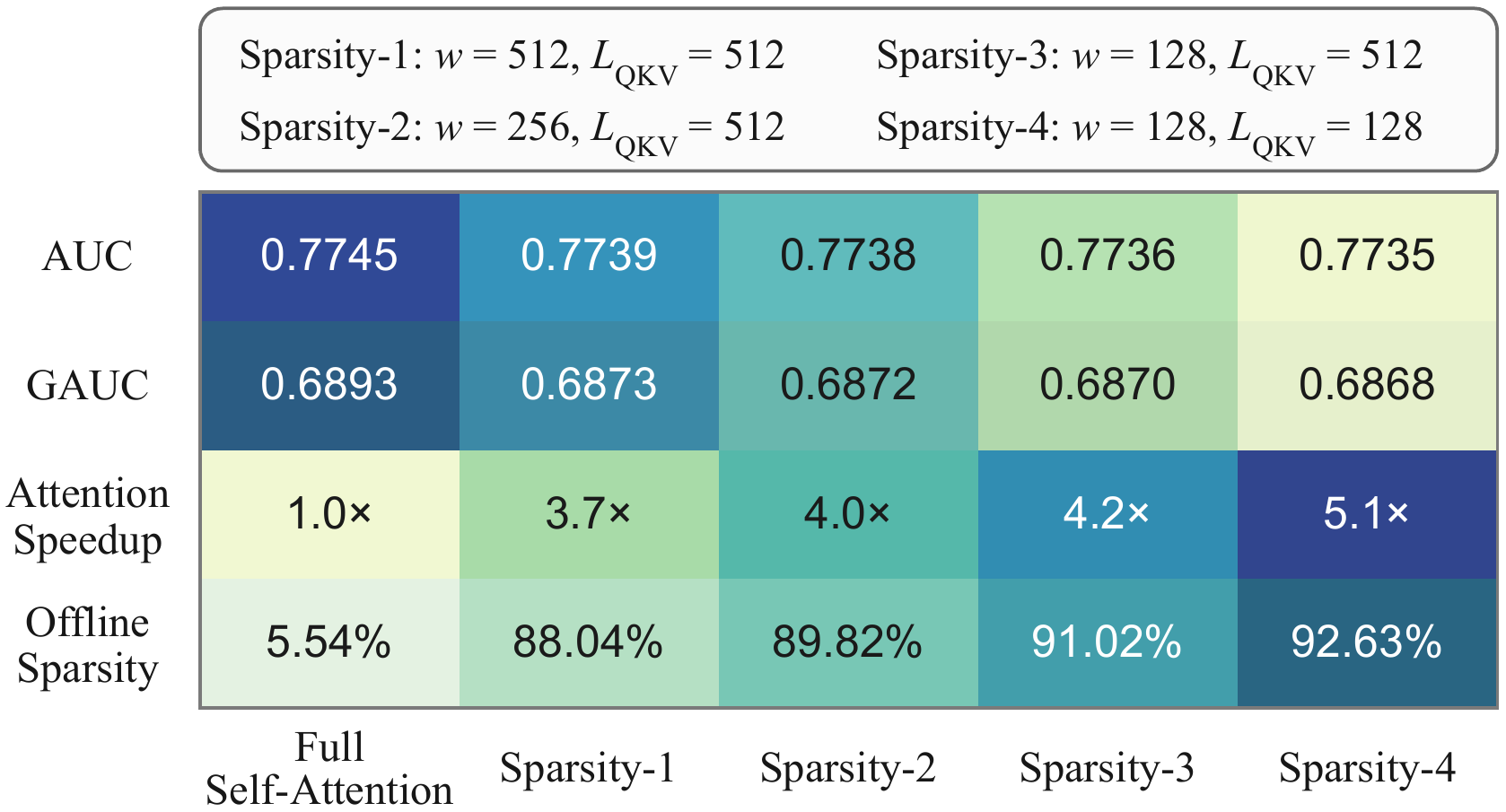}
\caption{Performance of hybrid sparse attention under different configurations in terms of CTR, speedup, and sparsity.}
\label{fig:sparse_attention}
\end{center}
\end{figure}

Next, we examine hybrid sparse attention by varying the sliding-window size $w$ and retained historical length $L_{\text{QKV}}$ after pruning to construct four configurations with different sparsity levels. The results are presented in Figure~\ref{fig:sparse_attention}.

We observe that all sparse configurations substantially accelerate attention computation while maintaining competitive CTR performance. Specifically, Sparsity-1 achieves a $3.7\times$ speedup with 88.04\% sparsity, while incurring only marginal AUC and GAUC degradation compared with full self-attention. As the sliding-window size decreases from Sparsity-1 to Sparsity-3, the sparsity and speedup continuously increase, whereas the performance declines only slightly. This indicates that attending to a limited local window is sufficient to capture most useful sequential dependencies.
Moreover, further pruning the retained historical representations in Sparsity-4 increases the sparsity to 92.63\% and achieves a $5.1\times$ attention speedup. We adopt Sparsity-4 because reallocating its saved compute to depth, sequence length, and epochs outweighs 0.10\,pt AUC cost, as shown in Figure~\ref{fig:scaling_laws}. These results demonstrate that the hybrid sparse attention proposed in LazFormer effectively removes redundant attention computation, 
achieving a favorable trade-off between recommendation performance and training efficiency.

\begin{table}[t]
\centering
\caption{Online A/B testing results.}
\label{tab:online_AB}
\setlength{\tabcolsep}{10pt}
\begin{tabular*}{0.47\textwidth}
{@{\extracolsep{\fill}}@{}lcccc@{}}
\toprule
\textbf{Metrics} & \textbf{IPV} &  \textbf{Orders} & \textbf{Buyers} & \textbf{GMV} \\
\midrule
\% Improve $\uparrow$ & +5.21\%  & +3.38\% & +3.70\% & +9.85\% \\
\bottomrule
\end{tabular*}
\end{table}

\subsection{Online A/B Testing (RQ6)}
\label{onlineab}

We deploy LazFormer in the ranking stage of the homepage recommendation system on a large-scale industrial e-commerce platform, where the online production baseline is a 3-layer SORT-like~\cite{sort} ranking model. A two-week online A/B test is conducted, and the results are reported in Table~\ref{tab:online_AB}. Compared with the baseline, LazFormer achieves relative improvements of 5.21\% in IPV (\textit{i.e.}, item page view), 3.38\% and 3.70\% in the numbers of orders and buyers, respectively, and 9.85\% in GMV. These consistent gains across engagement, conversion, and revenue metrics demonstrate the practical effectiveness and commercial value of LazFormer in real-world industrial recommendation.

Although LazFormer incurs additional online computation due to its deeper Transformer backbone (from 3 to 5 layers) and the modeling of longer sequences, the overhead is effectively controlled by its efficient sequence modeling design, including long-sequence compression and hybrid sparse attention. As a result, the single-machine throughput (measured in queries per second) on an NVIDIA A10 GPU decreases by only 12.1\%. Given the substantial business gains, LazFormer demonstrates strong practicality and a favorable return on investment for large-scale industrial recommendation.

\section{Related Work}

\textbf{Traditional industrial recommendation} models generally consist of two major components: user sequence modeling and feature interaction. Early sequence modeling methods~\cite{din,dien,bst,dsin} mainly focus on capturing users' evolving interests from relatively short behavior sequences. 
To incorporate longer behavior histories, subsequent studies~\cite{sim,eta,twin,transact} develop specialized retrieval and efficient encoding mechanisms. 
More recently, LONGER~\cite{longer} improves the scalability of long-sequence encoding through efficient attention, hierarchical aggregation, and serving-oriented optimization.
For feature interaction, early methods~\cite{wide_deep,deepfm,dcn,inttower,stem} capture implicit or explicit feature interactions through shallow neural networks. More recently, Wukong~\cite{wukong} and RankMixer~\cite{rankmixer,tokenmixer_large} explore scaling feature interaction networks by increasing model parameters and computational budgets. 
However, these methods cannot be expanded with a shared Transformer backbone, preventing them from fully exploiting the scaling capability of Transformer.


Inspired by the scaling capability of Transformer in LLMs, recent studies~\cite{hstu,mtgr,onetrans,mixformer,hyformer,tmallgs} have explored \textbf{scalable Transformer} architectures for industrial recommendation. HSTU~\cite{hstu} and MTGR~\cite{mtgr} reformulate recommendation as generative sequential modeling and demonstrate the potential of Transformers for large-scale recommendation. OneTrans~\cite{onetrans} employs a shared Transformer to unify feature interaction and sequence modeling. MixFormer~\cite{mixformer} and HyFormer~\cite{hyformer} further improve heterogeneous feature interaction and long-sequence modeling within scalable Transformer-based architectures. These studies demonstrate that recommendation performance can continuously benefit from scaling model depth, width, and sequence length.
However, most existing methods conduct scaling within a single ranking model and jointly optimize sparse and dense parameters from scratch, resulting in substantial computational resource consumption and slow convergence.
Recently, pre-training methods~\cite{gpsd,sort} introduce pre-trained sparse parameters into ranking, they typically freeze the sparse parameters, limiting their adaptation to the ranking process.
In contrast, LazFormer 
introduces a transferable residual adapter to mitigate the negative transfer of dense parameters and combines request-aware ranking with an asymmetric multi-epoch training strategy to enable the co-scaling of model capacity and training data while alleviating the overfitting of sparse parameters.

\section{Conclusion}

In this paper, we propose a scaling Transformer for industrial recommendation via transferable generative pre-training, named LazFormer. We first propose a generative pre-training module to autoregressively generate sequence features, providing favorable initialization for
both sparse and dense parameters in the subsequent ranking process. Additionally, we introduce a transferable residual adapter to incorporate ranking-specific features while preserving the pre-trained knowledge, thus solving the negative transfer of dense parameters. Moreover, the request-aware ranking module combines long-sequence compression, hybrid sparse attention, and a request-aware paradigm to efficiently model users' long interaction sequences. Besides, an asymmetric multi-epoch training strategy is developed to scale the training data by continuously accumulating dense parameters, while resetting sparse parameters to alleviate overfitting. Extensive offline experiments and online A/B testing demonstrate the effectiveness and scalability of LazFormer. 
LazFormer has been deployed in an industrial recommendation system, 
delivering significant gains for the platform.



\balance
\bibliographystyle{ACM-Reference-Format}
\bibliography{sample-base-extend.bib}

\end{document}